\documentclass[%
 reprint,
 amsmath,amssymb,
 aps,
]{revtex4-2}

\usepackage{graphicx}
\usepackage{dcolumn}
\usepackage{bm}

\begin{document}

\preprint{APS/123-QED}

\title{Influence of hydrometeors on relativistic runaway electron avalanches}

\author{D. Zemlianskaya}
 \affiliation{Moscow Institute of Physics and Technology - 1 “A” Kerchenskaya st., Moscow, 117303, Russian Federation}
 \affiliation{Institute for Nuclear Research of RAS - prospekt 60-letiya Oktyabrya 7a, Moscow 117312}

\author{E. Stadnichuk}%
\affiliation{%
HSE University  - 20 Myasnitskaya ulitsa, Moscow 101000 Russia\\ 
}%

\author{E. Svechnikova}
\affiliation{
 Institute of Applied Physics of RAS - 46 Ul'yanov str., 603950, Nizhny Novgorod, Russia%
}%

\date{\today}

\begin{abstract}
Previously, all studies in this area of atmospheric physics, namely, avalanches of relativistic runaway electrons (RREA), were carried out without taking into account the presence of hydrometeors in thunderclouds, which could seriously affect the results and their correspondence to actually observed natural phenomena.  This article takes into account hydrometeors in clouds. In this work, the distribution of RREA was simulated in GEANT4 was simulated taking into account various concentrations of ice particles. Modeling showed that accounting for the presence of hydrometeors cannot be simplified and reduced to a change in the main substance.
Two methods are considered - modeling of volumetric hydrometeors as separate modeling objects and as a simple change in the components of a whole substance (adding water to air with a corresponding density). These methods show completely different results.
Modeling by volumes of hydrometeors shows a decrease in the length of the avalanche by 20 \text{\%}, on the other hand, when modeling with a modified material, the length changed only by 1 \text{\%}. This suddenly proves that hydrometeors should be taken into account in research, as they can significantly change the growth length of an avalanche in real thunderstorm condition.

\end{abstract}


\maketitle

\section{Keypoints}
\begin{itemize}
\item The presence of hydrometeors contributes to the narrowing of the beam of avalanches of relativistic runaway electrons
\item The realistic density of hydrometeors in a cloud can reduce the avalanche growth length by 20 percent, which leads to the significant increase in the number of relativistic electrons in thunderstorms
\item The effect of runaway electrons multiplication is connected with the geometry of the hydrometeors, the contribution of their density and chemical composition proved to be neglectable
\end{itemize}

\section{\label{sec:level1}Introduction }

Research of high-energy phenomena in atmospheric electricity has been going on for several decades, but there are still many unsolved mysteries. One thing is known for sure --- the main participants in all high-energy thunderstorm processes are runaway electrons (\cite{fermi}, \cite{CHILINGARIAN2021102615}, \cite{asim} \cite{Gurevich_2001}).
Electrons in strong large-scale thunderstorm electric fields can obtain more energy from acceleration by the electric field than they in average lose on interactions with air molecules. Such accelerating electrons are called runaway electrons \cite{Gurevich1992, Dwyer2007}. Runaway electrons can produce additional runaway electrons by Moller scattering on air molecules \cite{Dwyer_2003_fundamental_limit}. In this way, runaway electrons multiply and form a relativistic runaway electron avalanche (RREA). Electric field strength necessary for RREA production is called critical electric field and depends on the air density \cite{Dwyer_2003_fundamental_limit}. RREAs are commonly studied using numerical calculations \cite{STADNICHUK2022106329} \cite{https://doi.org/10.1029/2021JD035278} \cite{Babich_2020} and Monte Carlo simulations (\cite{Dwyer_2012_RFDM}; \cite{https://doi.org/10.1029/1999JA900335}; \cite{https://doi.org/10.1002/2014JA020504}; \cite{refId0}; \cite{Khamitov_2020}). Analytical solutions for individual RREAs were described in (\cite{Gurevich_2001}; \cite{Babich:2020}; \cite{https://doi.org/10.1029/1999JA900335}).

Runaway electrons radiate bremsstrahlung gamma-rays when they interact with air. These gamma-rays are detected as high-energy component of TGE \cite{Chilingarian_2011_natural_accelerator} \cite{ Chilingarian_2020_radon}. Relativistic runaway electron avalanches are believed to cause thunderstorm gamma-ray glows \cite{Wada2019}. In addition, RREA bremsstrahlung is considered as one of possible sources of TGF \cite{asim_spectrum},  \cite{10_month_ASIM}, \cite{Fermi_2016}, \cite{Dwyer2012_phenomena}. TGF differs from the other high-energy atmospheric physics phenomena in its short duration and high brightness. For RREAs to cause a TGF large number of relativistic runaway electron avalanches is required \cite{Dwyer_vs_Gurevich}.

The main problem in the formation of a discharge in thunderclouds lies in the weak electric field, which is observed experimentally. The value of the observed differs from that required for a normal discharge by an order of magnitude (\cite{https://doi.org/10.1002/2015GL065620}, \cite{Stolzenburg2008}, \cite{Marshall1995}, \cite{Marshall1998_estimates}). Hence, there is a need to consider new effects that can increase the electric field and contribute to the formation of lightning and Thunderstorm Ground Enhancement (TGE).

However, it has never been possible to get enough electrons for lightning and for initiation TGF \cite{Dwyer_vs_Gurevich}, \cite{Gurevich_2001}, \cite{STADNICHUK2022106329}. With a large number of particles, we would observe a decrease in the required value of the electric field for breakdown due to mass character. Also, the strength of feedback would increase many times over \cite{STADNICHUK2022106329}, \cite{Dwyer2007}. The key parameter of an avalanche is its growth length, and for real conditions in a cloud, it is quite large, so there are few high energy particles in avalanche.

Naturally, a cell (a selected area of a thundercloud with a directed electric field \cite{Stadnichuk:2021ikv})  of a thundercloud cannot be considered ideally homogeneous. The mass density of hydrometeors in cumulonimbus clouds typically is up to 0.5 g/$m^3$ according to \cite{pruppacher1996microphysics}. Approximations are usually considered due to the small scale of the selected area. Moreover, the presence of hydrometeors should be taken into account. 

Previously, hydrometeors were considered in streamer physics. Studies have been carried out on the possible role of corona discharges on ice hydrometeors \cite{article1}. A class of hypotheses for the initiation of lightning and sprites suggests that streamers are able to form around the sharp tips of conducting objects (e.g., thundercloud hydrometeors for lightning and ionospheric ionization patches for sprites) placed in an electric field much weaker than the value of the electric field at which the electrons are accelerated rather than damped due to collision \cite{article2}. That is, impurities in clouds were previously considered as the cause of a change in the electric field, and not as an effect due to a change in the mass fraction of water in space, or, as the appearance of a dense inhomogeneity, a change in the direction of particle propagation and, as a consequence, a change in the total number of formed particles.

In this work, we have shown that unexpectedly hydrometeors efficiently multiply runaway electrons. By analogy with the interaction with air particles, runaway electrons can interact with hydrometeors in thunderclouds. This results in a 20\text{\%} reduction in avalanche growth length for a realistic number of hydrometeors and thus in the increase of the number of electrons.

In this article, the influence of the presence of hydrometeors in thunderclouds is studied. Also, the influence of the atomic composition of the substance is considered specifically, without taking into account the change in the electric field by the particles of hydrometeors. This article presents the analysis of RREAs development in the cloud, based on numerical modeling. Describes two different modeling options taking into account hydrometeors. The second section "Modeling" describes the modeling of energetic particles behaviour in the media of the cloud modeled in two different ways: model with volumetric hydrometeors and model with modified material. Section 2.2 presents the simulation results with volumetric hydrometeors. A comparison is made of the number of produced electrons with an increase in the mass fraction of hydrometeors and one radius, as well as one mass fraction and different radius.A comparison of the spectrum at the exit of a cell with and without a small number of volumetric hydrometeors, and a simulation with one hydrometeor is also presented. Section 2.3 presents the simulation results with modified material. The results in the section prove that the effect obtained is not a consequence of a change in the nuclear composition of the substance (adding an ice component).

\section{Modeling}

\subsection{General properties}

Numerical modeling is one of main instruments in atmospheric physics, crucial in conditions of limited measurement data, and useful for analyzing complex mechanisms verifying the results of other studies.
In present study we carried out Monte Carlo simulation using GEANT4 version 4.10.06.p01. GEANT4 is widely known and used for the problems of simulating the passage of particles through matter. Its fields of application include high energy physics, nuclear and accelerator physics, as well as medical and space research \cite{geant4}.

We used the Physics list \texttt{G4EmStandardPhysics\_option4}. 
It should give accurate results in electromagnetic physics simulations.
We developed two approaches to modeling the cloud media with hydrometeors: model with volumetric hydrometeors and model with modified material.
Within both approaches, the modeling volume was a cube with a side of 200~m. The cube initially consists of air with a density of 0.414 mg/c$m^3$, which corresponds to the density of air at a height of 10 km. Next, depending on the chosen approach, we add hydrometeors. In the cube the uniform electric field with strength 200 kV/m is applied. The direction is chosen so that the electrons are accelerated in the direction of launch. The minimum step for simulation was chosen as fMinStep = 0.01 mm.  An electron with an energy of 5 MeV is launched. We fix the born particles in the entire volume using the SensitiveDetector (the sensitive detector class in Geant4 has the task of creating hits (deposits of energy) each time a track traverses a sensitive volume and loses some energy).

\subsection{Model with volumetric hydrometeors}

In this simulation, hydrometeors with a radius of 1 cm were used, which is bigger than a real hydrometeor. However, modeling smaller balls leads to an increase in their number at the same mass fraction, which complicates the calculations. In order to justify the use of particles of this size, the following additional simulations were carried out.

\begin{figure}[h!]
    \centering
    \includegraphics[width=0.7\linewidth]{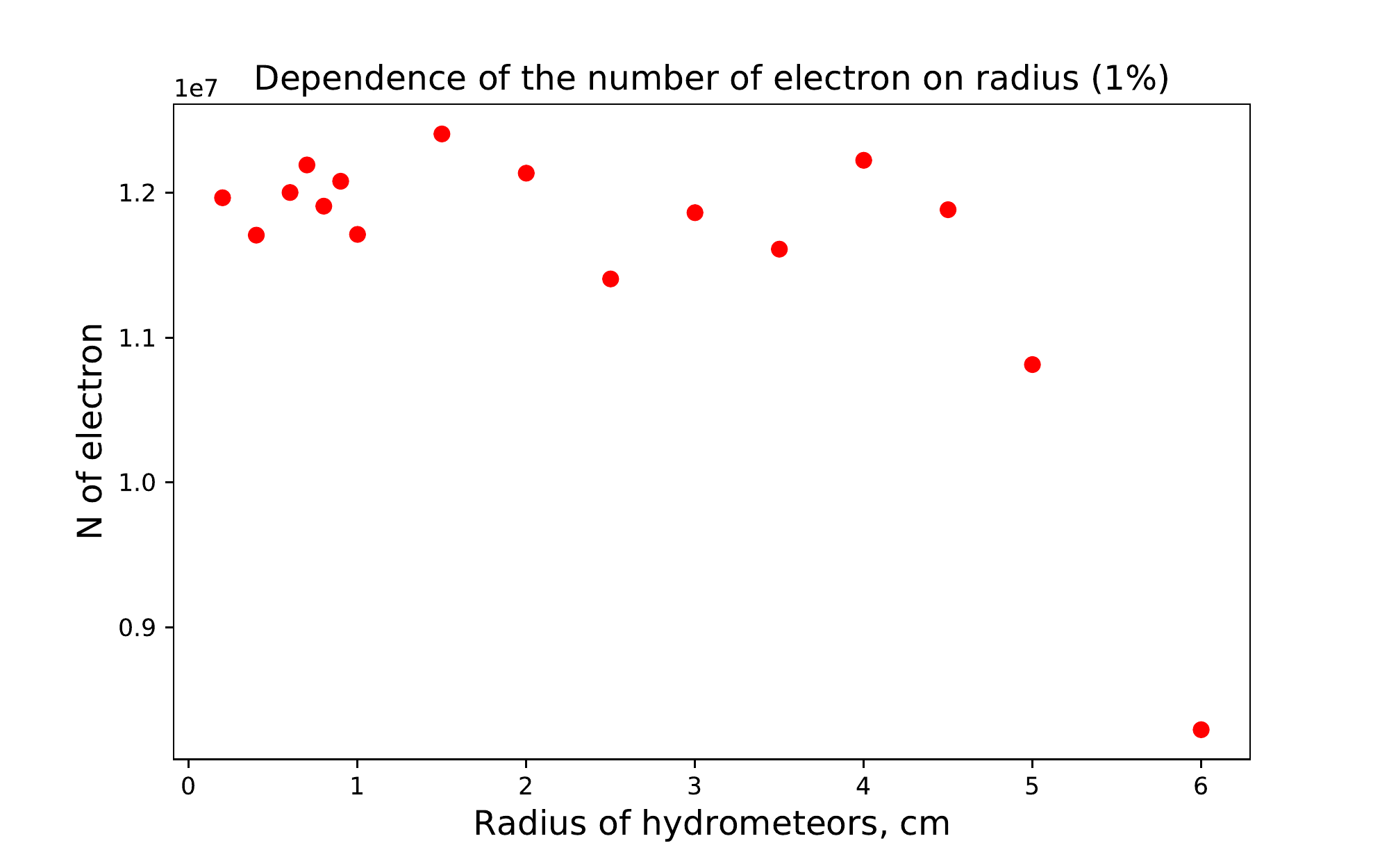}
    \caption{Dependence of the number of electrons on the radius of the formed hydrometeor. Modeling at an air density of 0.414 mg/cm3 and field 200 kV/m. As you can see, the size has little to do with it.}
    \label{fig:snow_electrons2}
\end{figure}

A cube with a side of 50 meters was taken, since in this case one can limit oneself to measuring the number of electrons, and not the growth length of the avalanche. For exact values of the growth length, one should take the length of several the growth length of an avalanche so that it has time to stabilize. For us, in order to model many small-radius hydrometeoroids for a given mass fraction, we need to minimize unnecessary calculations, therefore we will consider the number of electrons in a small cell. A thousand electrons are also launched from one side of the cube. Hydrometeors are scattered in the cube itself. Having fixed the mass density of hydrometeors in the cell at the level of 1 percent and changing the particle radius, we obtain the following dependence of the number of generated electrons on the radius \ref{fig:snow_electrons2}.

The first modeling approach allows us to analyze how the presence of non-homogeneous material, namely a sharp contrast in density between hydrometeor and air volumes, affects electron avalanches.
To study the material-caused influence, we place balls with a radius of 1 cm in a randomly evenly in a cube with a side of 200 m. Density of hydrometeors 900 mg/cm3.
By changing the number of hydrometeors, we obtain the distributions of produced electrons in the entire volume, and with it the growth lengths.

\begin{figure}[h!]
    \centering
    \includegraphics[width=0.7\linewidth]{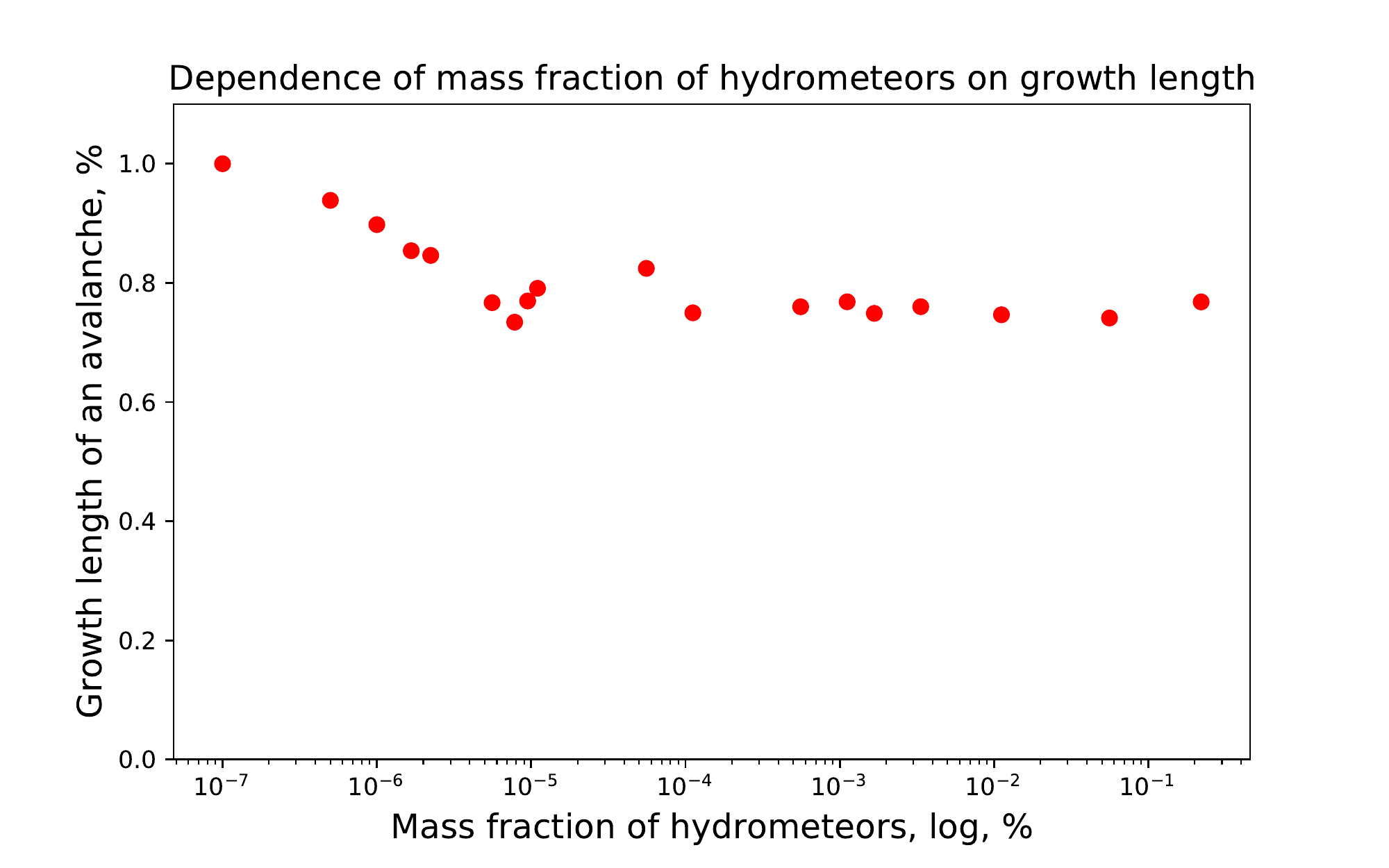}
    \caption{Dependence of mass fraction of hydrometeors on the growth length of an avalanche.  Hydrometeors are balls with a radius of 1 cm. Modeling at an air density of 0.414 mg/cm3 and field 200 kV/m}
    \label{fig:snow_electrons1}
\end{figure}

As one can see in the Fig.\ref{fig:snow_electrons1}, with an increase in the number of hydrometeors in the considered volume, the number of generated electrons increases. Presumably this is a consequence of the fact that as the electrons move, they lose energy interacting with hydrometeors and producing new particles, which are also accelerated by the electric field and undergo collisions. Since there are not so many hydrometeors, after each collision generated particles can gain enough energy for generated particles in a next collision and run away, and then produce additional runaway electrons in further collisions.

In order to be convinced of the influence of a small number of hydrometeors on the scale of the entire avalanche, one should look at interactions with 1 hydrometeor. In the next simulation a beam of 500 keV was launched on one hydrometeor with different sizes. Simulated volume is a cell with a side of 50 cm. 1 hydrometeor is placed in its center. The field is also 200 kV/m. In the case of a hydrometeor with a radius of 1 cm - 473939, 1mm - 478136 particles. In the same simulation, but without the hydrometeor, 445107 particles are born. The difference is significant. Hydrometeors increase the number of runaway electrons, which is why a significant difference is visible on the scale of the entire avalanche.

\begin{figure}[h!]
    \centering
    \includegraphics[width=0.7\linewidth]{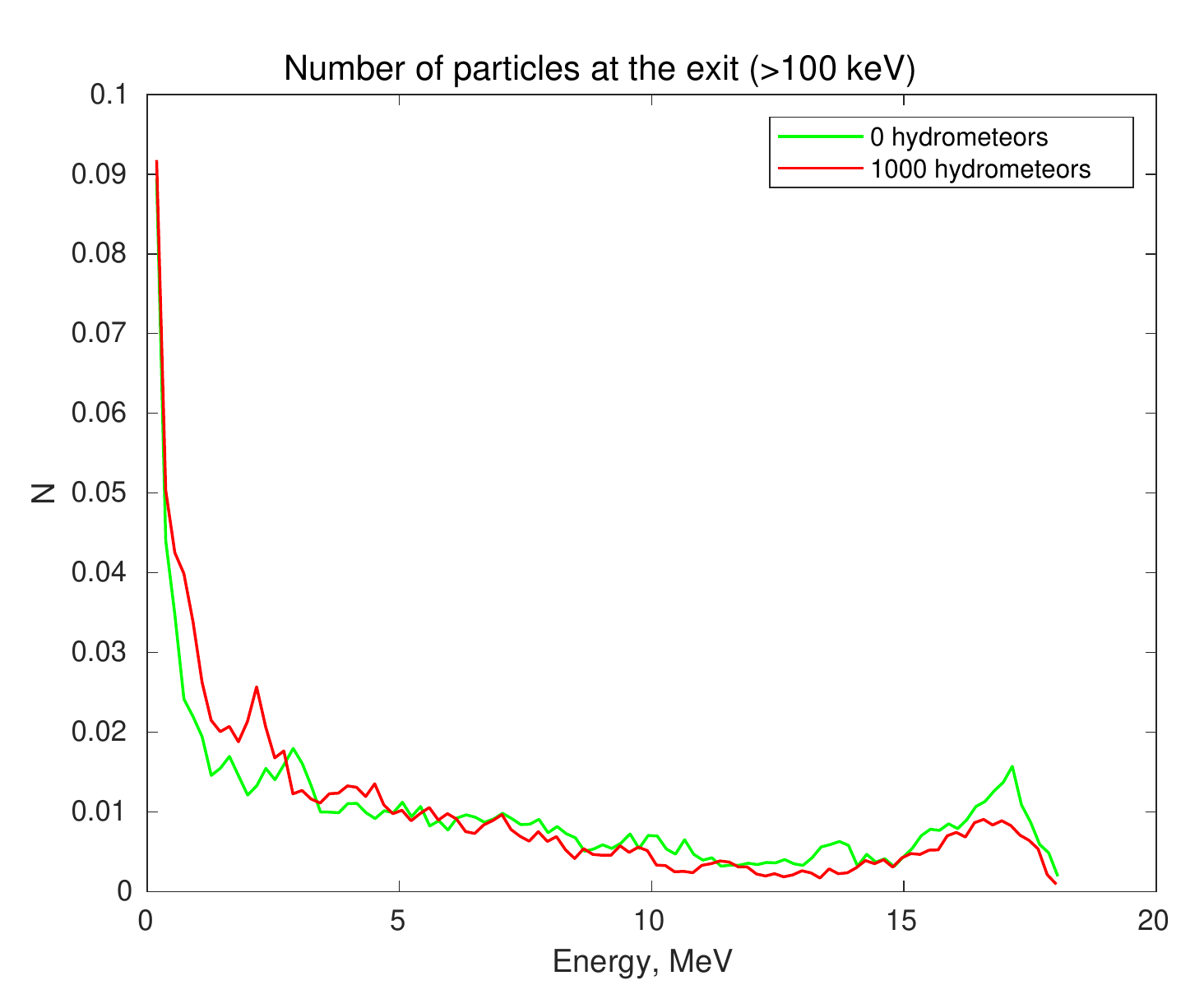}
    \caption{Spectrum of produced particles with energy above 100 keV at exit a cell with a side of 200 m.  Green line - 0 hydrometeor. Red line - 1000 hydrometeors. Modeling at an air density of 0.4~mg/cm3 and field 200~kV/m}
    \label{fig:snow_spectr}
\end{figure}

The simulated volume is a cube with a side of 200 m. In the volume of it we place 1000 hydrometeors with a radius of 1 cm. As previously clarified for 1 hydrometeor, this size does not change the mechanism for increasing the number of particles. It is also worth mentioning that, as in the leading simulations, the field is considered equal to 200 kV/m, the air density in the cube is 0.414 mg/cm3, the  density of the hydrometeor is 900 mg/cm3. From the point -50 m along the z-axis (the volume is located from -100 to 100 m), electrons with an energy of 5 MeV are launched in the direction of electron acceleration by the electric field.
An air detector is placed at the exit from the specified volume - a volume of air of the same density with a layer width of 10 m. The generated particles are fixed precisely in this layer after leaving the cube. Also, the obtained data was filtered by the detected particle energy: only particles above 100 keV are detected, since only particles with energies above can be accelerated by the electric field and become runaway electrons, thereby the avalanche growth length. In the absence of hydrometeors, 41047 particles were obtained, with the presence of 1000 hydrometeors - 58326 particles. Also, the spectra obtained as a result of the simulation normalized to the total number of produced particles are shown in \ref{fig:snow_spectr}. The figure shows that the spectrum does not change, but the number of particles changes. The spectrum is interrupted at energy 17 MeV because the cell is short and the particles leave it, while in reality it extends up to the potential difference of the cell.

\subsection{Model with modified material}

Due to hydrometeors, new runaway electrons appear, but also due to their high density compared to air, they can lead to particle damping \cite{Babich_2020}. It is also worth bearing in mind that we change the atomic composition of the substance, because water is added. Next, simulations will be carried out to investigate the impacts of these features.

Hydrometeors typically have size 1 nm – 100 $\mu$m, which on average is smaller than that in the simulation. Reducing the size of each hydrometeor while maintaining their mass fraction in the system will inevitably lead to an increase in the number of required volumes. This will significantly increase the load on the device and the simulation time. To avoid an infinite number of hours of data collection, a simplified model can be considered. Assuming that the particles are only dust when considering a large volume, it is possible to include their properties in the common material in the air. Technically, GEANT4 will not create multiple volumes, but only one, to which the new material will be assigned. Thus, it will greatly facilitate the acquisition of data in a system with hydrometeors. This will allow you to reach the largest possible number of hydrometeors in the system by specifying a material consisting entirely of water with an overdetermined density.

In order to test this assumption about the possibility of using only one own material from a mixture of air and ice instead of modeling many volumes, a GEANT4 simulation was carried out. 
It is a cube with sides of 200 meters, consisting of a new material ``Mix air snow''. This material is a mixture consisting of two components --- G4WATER with a redefined density of 900 kg/m3 and G4AIR with 0.414 mg/cm3. A feature of setting the material in this way is the indication of the total density of the volume. Two cases can be considered. The first is that the density of the common material remains unchanged, but the mass ratio of two types of media is changed. Second way to define the media is to change the density of uniform media, depending on the proportion of incoming substances.

In both cases, in order to see in this way the dependence of the number of particles on the number of hydrometeors, the ratio of components in the mix air snow material was changed in the simulations. The ratio of hydrometeors in the new material is indicated on the x-axis in the presented figures.

\begin{figure}[h!]
    \centering
    \includegraphics[width=0.7\linewidth]{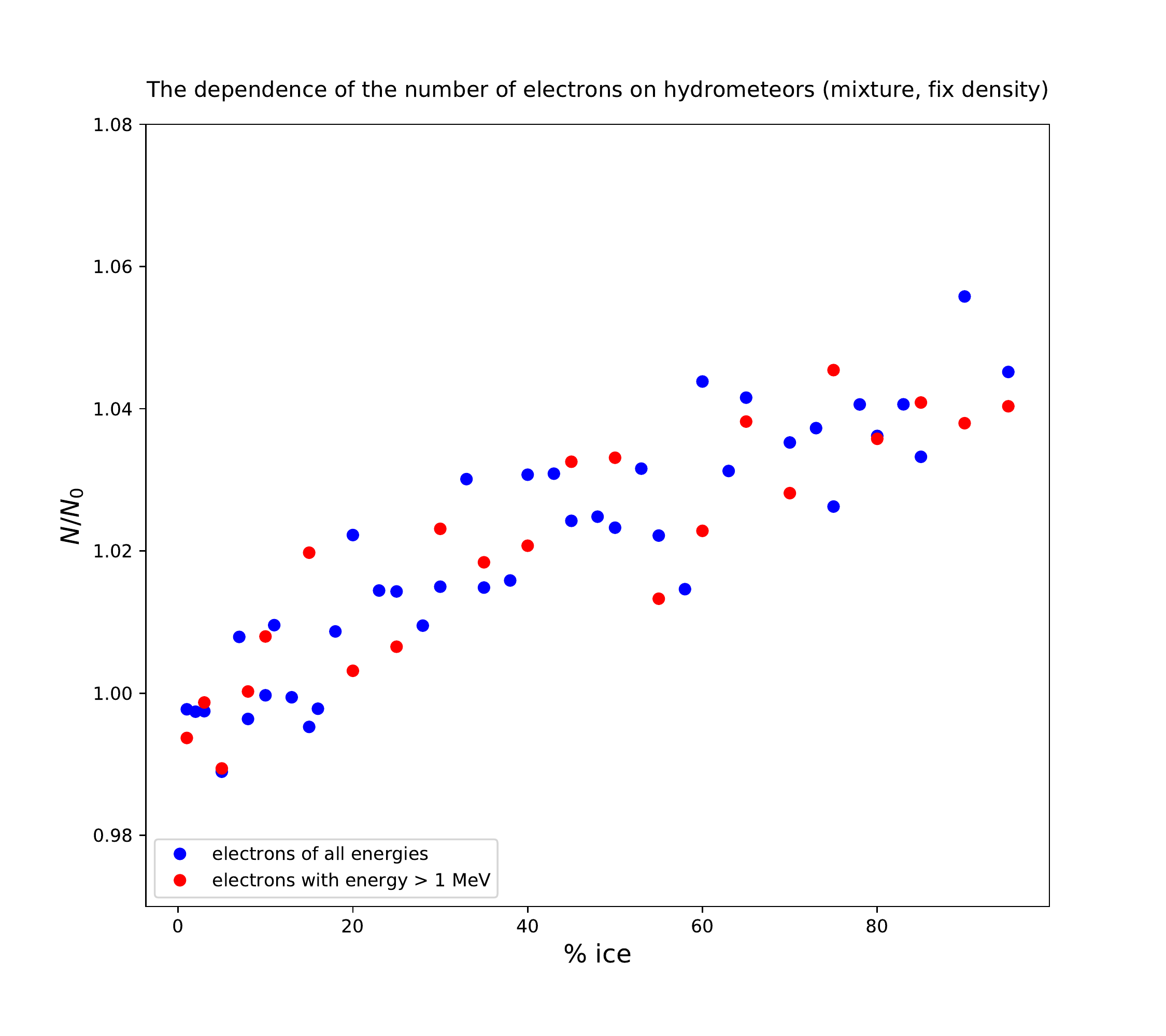}
    \caption{Dependence of the number of electrons on the mass ratio ice in the material. Simulation at air density 0.414 mg/cm3 and field 200 kV/m. On the figure \ref{fig:snow_electrons1}, electrons are generated at birth in the volume itself.}
    \label{fig:mix_snow_electrons}
\end{figure}

\begin{figure}[h!]
    \centering
    \includegraphics[width=0.6\linewidth]{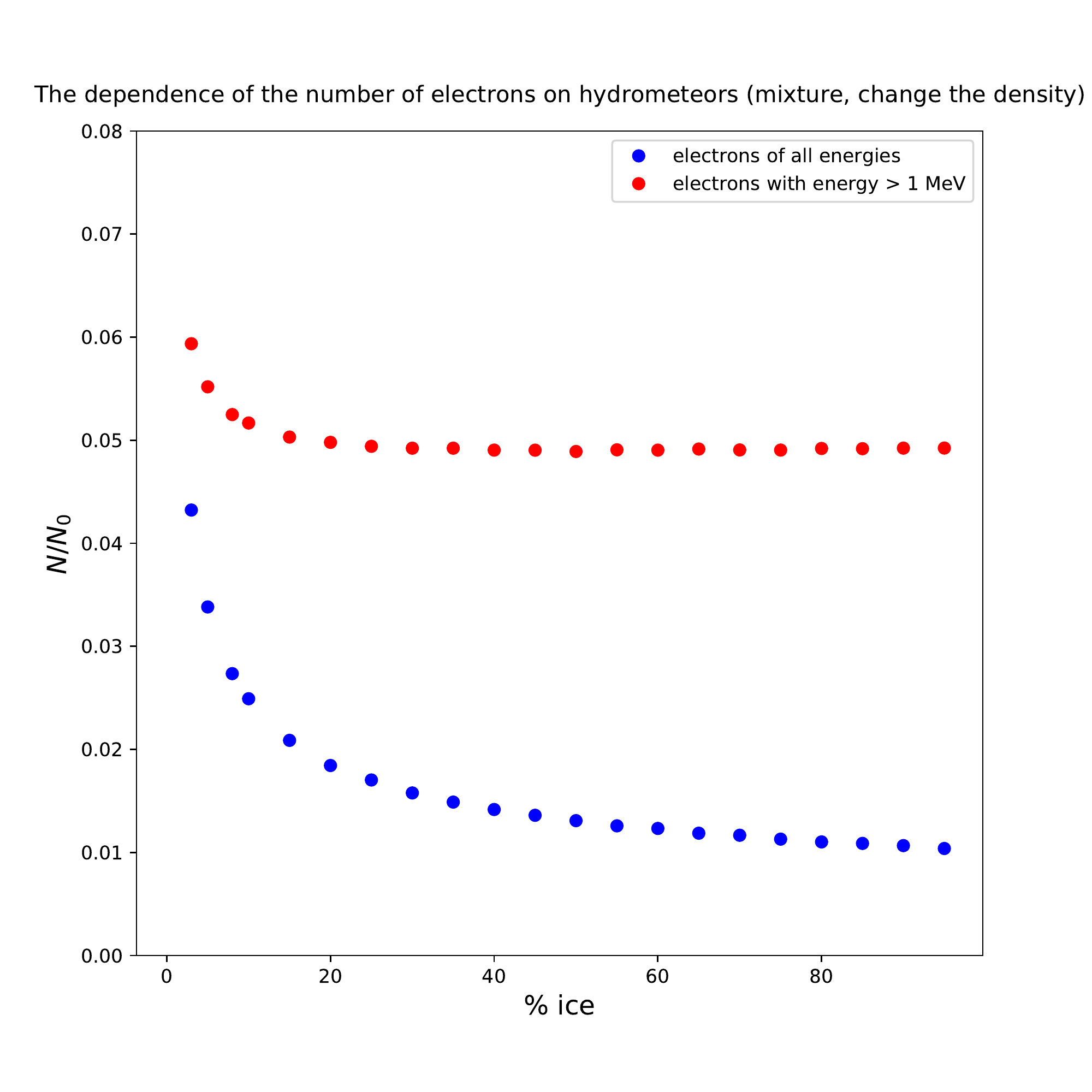}
    \caption{Dependence of the number of electrons depending on the percentage of the mass fraction of ice in the material. In this case, hydrometeors are included as a change in the material components of the total volume. The total density of the substance varies depending on the percentage of occurrence of each of the components. Simulation at air density 0.4 mg/cm3 and field 200 kV/m}
    \label{fig:mix_snow_electrons_change_den}
\end{figure}

The resulting figure are \ref{fig:mix_snow_electrons}, \ref{fig:mix_snow_electrons_change_den}. The first figure shows that with a small proportion of hydrometeors, the number of electrons decreases by 1 percent. With an increase in the proportion of water in the material over 20 percent, we see an increase in the number of particles by fractions of a percent. This is not at all equivalent to the results of the previous section on volumetric hydrometeors. Since in this simulation the total density remains unchanged, only the change in the atomic composition of the substance plays. The total effect obtained in the previous section cannot be explained by a change in the atomic composition of the substance. In the second figure, the total density changes in the same ratio as the proportion of water in the material changes. 
It is shown that with an increase in the number of hydrometeors, particle attenuation occurs, which is logical. However, in a realistic case, the mass fraction of hydrometeors is up to 5 g/$m^3$, or up to 1.25 percent. It follows from this that this method of hydrometeor modeling cannot show the same result as volumetric hydrometeor modeling, due to its uniformity and increasing density.

\section{Discussion}
It turns out that hydrometeors affect avalanches of relativistic runaway electrons, significantly. Hydrometeors multiply the number of runaway electrons very efficiently. 
Although it would seem that due to the increased density of the entire material, electrons should, on the contrary, be less (\cite{Dwyer_2003_fundamental_limit}, \cite{Babich_2020}, \cite{Gurevich1992}). According to \cite{Dwyer2007}, The avalanche (e-folding) length is well fit by the empirical relation:
$$\lambda = \frac{7300 kV}{(E - (276 kV/m) n/ n_0)}$$ valid over the range 300–2600 kV/m 300-2600, where the electric field, E, is measured in kV/m \cite{Dwyer_2003_fundamental_limit} \cite{https://doi.org/10.1029/2006GL025863}. However, now this formula should also include a correction for hydrometeors in clouds, which affects the entire studied physics of thunderclouds.

It was shown that the very effect of density, as well as changes in the nuclear composition of the substance (adding ice components) of the hydrometeor, has very little effect on the formation of an avalanche - about a percent \ref{fig:mix_snow_electrons}.

In the case of bulk snowflakes, the runaway electrons in them multiply greatly, and the runaway electrons born in the collision have enough distance between the snowflakes to accelerate to an energy of several MeV and, when colliding with another snowflake, multiply strongly. This leads to the observed strong enhancement of avalanches in the system with hydrometeors \ref{fig:snow_electrons1}.

It has been established that with an increase in the number of hydrometeors in a thundercloud, the number of electrons produced in a given volume under the action of an electric field increases, if we consider small cells of the order of 2 avalanche growth lengths. A strong dependence - a change in the  growth length up to 20 \text{\%} \ref{fig:snow_electrons1} - is observed precisely when considering a small mass fraction of volumetric hydrometeors. That is, at fairly realistic values - up to 1 \text{\%} of the mass fraction of ice in a thundercloud. It is possible that if there are too many hydrometeors, then, naturally, the density will kill the electrons. But in the case of thunderclouds, there are just enough hydrometeors to increase the number of runaway electrons.

The very reason for such a serious change in beam propagation is hidden in the geometry. The observed effect is explained by the fact that a small inclusion of dense particles creates points for electron multiplication. Due to the small number of such points, this does not affect the overall density. There is also a lot of space between these points so that the electrons do not decay.

\begin{figure}[h!]
    \centering
    \includegraphics[width=0.8\linewidth]{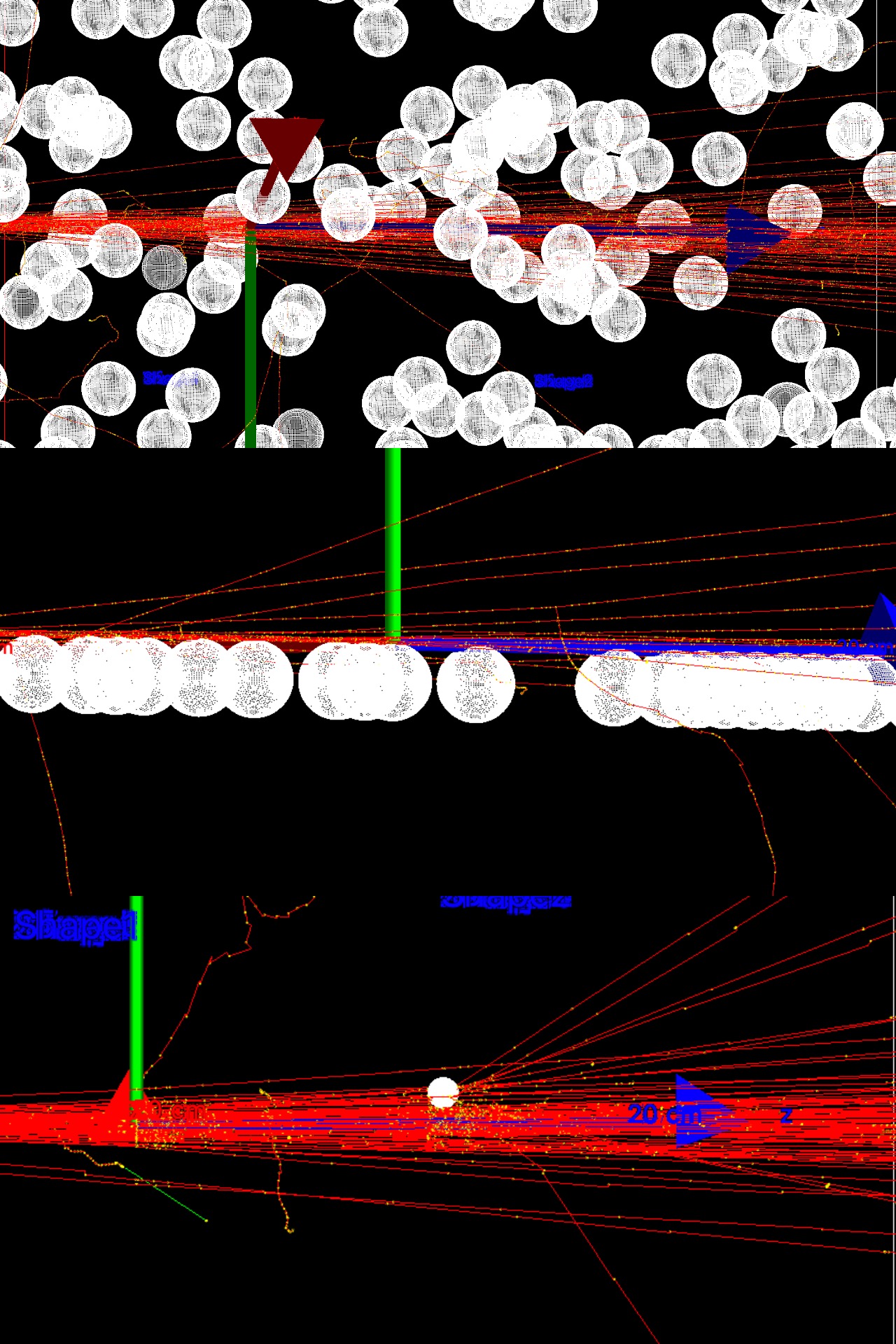}
    \caption{Visualization of beam interaction with hydrometeors. Simulation in Geant4. Red line - tracks of electrons. The top picture shows the propagation of the beam through many hydrometeors. The second picture illustrates the repulsion of the beam by hydrometeors. The last picture shows the interaction of the beam with only one hydrometeor.}
    \label{fig:model_geant}
\end{figure}

Also \ref{fig:model_geant} shows how electrons propagate among hydrometeors.
In addition, simulations and visualization \ref{fig:model_geant} show that the beam narrows in the presence of hydrometeors. This is important for the gamma radiation pattern \cite{Angular_DistributionTGF}, important for transverse diffusion in feedback \cite{STADNICHUK2022106329}. 

The number of particles increases and this is important for TGF and TGE (\cite{Babich_2020}), and it is important for feedback (\cite{Dwyer2007},  \cite{STADNICHUK2022106329}, \cite{https://doi.org/10.1029/2021JD035278}), since it is the runaway electrons that play the primary role.

This fact suggests that modeling and calculations should take into account the presence of hydrometeors, namely, the presence of separate volumes in them, which cannot be replaced by changing the material of the modeling volume. The study also shows that we can accurately use hydrometeors with a radius of up to 4 centimeters in simulations, consider in \ref{fig:snow_electrons2} even though real hydrometeors are much smaller.

It should be noted that taking into account hydrometeors when modeling the multicell reactor \cite{Stadnichuk:2021ikv}, relativistic feedback discharge model \cite{Dwyer2012_phenomena} due to the simpler occurrence of avalanches, it is cat to achieve a decrease in the values of the electric field value required for self-sustainable RREA development due to feedback processes. The opposite effect is also possible, since it will be more difficult for particles to turn around and form a mutual bond due to hydrometeors in the volume.
Evidently the influence of hydrometeors on feedback effects in s should be among the subjects of further research. 

An increase in the number of electrons affects primarily the number of particles in RREA. They, in turn, are a fundamental phenomenon for most of the processes that are observed in atmospheric high-energy physics. It is also important for lightning initiation \cite{https://doi.org/10.1002/2015GL065620}. This is why the found effect is very interesting.

Naturally, we should not forget that in this work we do not consider changes in the electric field that arise due to the presence of hydrometeors. This effect is planned to be taken into account in future research. However, in the first approximation, it is superfluous, because a small local field has almost no effect on runaway electrons and it complicates finding and explaining the effect caused by the inclusion of high-density particles on runaway electrons.

Summing up, it should also be noted that this work is an approximation of a conventional homogeneous cell, which is considered in most simulations, to more realistic conditions due to the presence of a small number of hydrometeors. The mass density in cumulonimbus clouds is typically up to 0.5 g/$m^3$ according to \cite{pruppacher1996microphysics}. When describing especially dense clouds and local inhomogeneities, 5 g/$m^3$ can be used as the upper limit of the cloud density, which corresponds to 1.25\text{\%} of the snow mass. This also applies to the quantities considered in this article, which once again confirms the need to include hydrometeors in the models under study.

\section{Conclusions}
It was shown that the influence of hydrometeors is significant for the propagation of runaway electrons in a cloud. Electrons colliding with hydrometeors multiply and are further accelerated by the electric field. Inclusions of high-density particles slightly affect the density of the medium, but radically change the beam propagation only due to its geometry. Therefore, the density of hydrometeors directly affects the development of avalanches of runaway electrons. In total, a small amount of hydrometeors (50 hydrometeors $\approx$ 0.19 kg in cube with a side of 200 meters $\approx$ $2.35*10^{-8}$ $kg/m^3$) can lead to a decrease in the growth length of an avalanche by 20 \text{\%}. This significantly changes the understanding of physics in RREA.

\nocite{*}

\bibliography{apssamp}

\end{document}